\documentclass[letterpaper, 10 pt, conference]{ieeeconf}
\IEEEoverridecommandlockouts
\usepackage[cmex10]{amsmath}
\usepackage{array}
\usepackage{url}
\usepackage[utf8]{inputenc}  
\usepackage{citesort}
\usepackage[T1]{fontenc} 
\usepackage{mathtools}
\usepackage{amssymb,amsfonts,amsmath}
\usepackage{dsfont}
\usepackage{hyperref}
\usepackage{dsfont}
\usepackage{stmaryrd}
\usepackage{bm}
\usepackage{flushend}

\def \({\left(}
\def \){\right)}
\def \[{\left[}
\def \]{\right]}
\newcommand{\argmin}{\text{argmin}}
\newcommand{\tbf}[1]{{\textbf{#1}}}

\newcommand{\defeq}{\vcentcolon=}

\newcommand{\bY}{{\textbf {Y}}}

\newcommand{\bZ}{{\textbf {Z}}}

\newcommand{\bX}{{\textbf {X}}}
\newcommand{\bx}{{\textbf {x}}}

\newcommand{\by}{{\textbf {y}}}
\newcommand{\bz}{{\textbf {z}}}

\newcommand{\bs}{{\textbf {s}}}
\newcommand{\bS}{{\textbf {S}}}

\newcommand{\be}{\begin{equation}}
\newcommand{\ee}{\end{equation}}
\newcommand{\bea}{\begin{eqnarray}}
\newcommand{\eea}{\end{eqnarray}}

\newcommand\smallO{
  \mathchoice
    {{\scriptstyle\mathcal{O}}}
    {{\scriptstyle\mathcal{O}}}
    {{\scriptscriptstyle\mathcal{O}}}
    {\scalebox{.7}{$\scriptscriptstyle\mathcal{O}$}}
  }

\newtheorem{theorem}{Theorem}[section]
\newtheorem{lemma}[theorem]{\textbf{Lemma}}
\newtheorem{thm}[theorem]{\textbf{Theorem}}
\newtheorem{remark}[theorem]{\textbf{Remark}}

\newtheorem{corollary}[theorem]{\textbf{Corollary}}
\newtheorem{definition}[theorem]{\textbf{Definition}}

\DeclareMathAlphabet{\varmathbb}{U}{bbold}{m}{n}

\newcommand{\EE}{\mathbb{E}}

\title{\LARGE \bf The Mutual Information in Random Linear Estimation}
\author{Jean Barbier$^{\dagger}$, Mohamad Dia$^{\dagger}$, Nicolas Macris$^{\dagger}$ and Florent Krzakala$^*$\\
$\dagger$ Laboratoire de Th\'eorie des Communications, Facult\'e Informatique et Communications,\\
Ecole Polytechnique F\'ed\'erale de Lausanne, 1015, Suisse. \\
$*$ Laboratoire de Physique Statistique, CNRS, PSL Universit\'es \&
Ecole Normale Sup\'erieure, \\Sorbonne Universit\'es et Universit\'e Pierre \& Marie Curie, 75005, Paris, France.}
\begin{document}
\maketitle
\begin{abstract}
We consider the estimation of a signal from the knowledge of its noisy linear random Gaussian projections, a problem relevant in compressed sensing, sparse superposition codes or code division multiple access just to cite few. There has been a number of works considering the mutual information for this problem using the heuristic replica method from statistical physics. Here we put these considerations on a firm rigorous basis. First, we show, using a Guerra-type interpolation, that the replica formula yields an upper bound to the exact mutual information. Secondly, for many relevant practical cases, we present a converse lower bound via a method that uses spatial coupling, state evolution analysis and the I-MMSE theorem. This yields, in particular, a single letter formula for the mutual information and the minimal-mean-square error for random Gaussian linear estimation of all discrete bounded signals.
\end{abstract}
Random linear projections and random matrices are ubiquitous in
computer science, playing an important role in machine
learning~\cite{johnson1984extensions},
statistics~\cite{mehta2004random} and
communication~\cite{tulino2004random}. In particular, the task of estimating a signal from its linear random projections has a myriad of
applications such as compressed sensing (CS)~\cite{candes2006near}, code
division multiple access (CDMA) in
communication~\cite{verdu1999spectral}, error correction via sparse superposition codes~\cite{barron2010sparse}, or Boolean group testing~\cite{atia2012boolean}. It is thus natural to ask what are the
 information theoretic limits for the estimation of a
signal via the knowledge of few of its (noisy) random linear
projections.

A particularly influential approach to this question has been through
the use of the heuristic replica method of statistical physics~\cite{mezard1990spin}, which allows to compute non rigorously the
mutual information (MI) and the associated theoretically achievable minimal-mean-square error (MMSE). The replica method typically predicts the optimal performance through the solution 
of non-linear equations,
which interestingly coincide in many cases with the predictions for the performance of a message-passing belief-propagation type algorithm. 
In this context the algorithm is usually called approximate 
message-passing (AMP)~\cite{donoho2009message,krzakala2012statistical,krzakala2012probabilistic}. 

In this contribution we prove rigorously that the replica formula for the MI is asymptotically exact for discrete bounded prior 
distributions of the signal, in the case of random Gaussian linear projections. In particular, our results put on a firm 
rigorous basis the Tanaka formula for CDMA~\cite{tanaka2002statistical}, and allow to rigorously obtain the Bayesian 
``measurement'' MMSE in CS. In addition, our analysis strongly suggests that AMP is reaching the MMSE for a large 
class of such problems in polynomial time, except for a region called the \emph{hard phase}. In the hard phase 
the MMSE can be reached only through the use of a technique 
called \emph{spatial coupling}~\cite{donoho2013information,krzakala2012statistical,krzakala2012probabilistic} (SC), originally developed 
in the context of communication as a practical code construction that allows to reach the 
Shannon capacity~\cite{Felstrom1999}. Finally, we stress that our proof technique has an interest of its own as it is probably transposable to various inference problems.

The paper is organized as follows. In sec.~\ref{sec:partI}, we introduce the problem and our notations, discuss the related previous works, introduce AMP and state evolution, present our main results and elaborate on the possible scenarios covered 
by our proof. Sec.~\ref{sec:partII} presents our proof strategy and sec.~\ref{sec:partIII} sketches the proofs of the main technical propositions.
\section{Setting, results and related works}
\label{sec:partI}
\subsection{Linear estimation: setting and summary of results}
\label{sec:settings}
In Gaussian random linear estimation, one is interested in reconstructing a signal $\bs\!\in \!\mathbb{R}^N$ from few 
measurements $\by\!\in \!\mathbb{R}^M$ obtained from a random i.i.d Gaussian \emph{measurement matrix} $\bm{\phi}\!\in\! \mathbb{R}^{M \times N}$,
\be \label{eq:CSmodel}
\tbf{y} = \bm{\phi}\tbf{s} + \tbf{z} \sqrt{\Delta}  \ \Leftrightarrow \
y_{\mu} = \sum_{i=1}^N\phi_{\mu i} s_i + z_\mu \sqrt{\Delta},
\ee
where the additive white Gaussian noise (AWGN) of variance $\Delta$ is i.i.d with $Z_\mu\! \sim \!\mathcal{N}(0,1)$, $\mu\!\in\!\{1,\dots, M\}$. The signal $\bs$ to be reconstructed is made of $L$ i.i.d \emph{sections} $\bs_l\! \in\! \mathbb{R}^B, l\!\in\!\{1,\ldots,L\}$, distributed according to a discrete prior $P_0(\bs_l) \!= \!\sum_{i} p_i \delta(\bs_l \!-\! \tbf{a}_i)$ with a finite number of terms and all $\tbf{a}_i$'s bounded. We will refer to such priors simply as 
{\it discrete priors}. Thus the total number of signal components is $N\!=\!LB$. The case of priors that are mixtures of discrete and 
absolutely continuous parts can presumably be treated in the present framework but this
leads to extra technical complications. The matrix $\bm{\phi}$ has i.i.d Gaussian entries $\phi_{\mu i}\!\sim\!\mathcal{N}(0,1/L)$. The \emph{measurement rate} is $\alpha\!\defeq\! M/N$. Equation \eqref{eq:CSmodel} is referred as the \emph{CS model} despite being more general 
than CS, and we borrow vocabulary of this field.

Define $\bar \bx \!\defeq\! \bx \!-\! \bs$, $[\bm{\phi}\bar \bx]_\mu\! \defeq \!\sum_{i=1}^N \phi_{\mu i} \bar x_i$. In the Bayesian setting, the posterior associated with the CS model is 
\begin{align*}
P^{\rm cs}(\bx|\by) &= \frac{\exp\Big(-\frac{1}{2\Delta}\sum\limits_{\mu=1}^M ([\bm{\phi}\bar \bx]_\mu - z_\mu \sqrt{\Delta})^2\Big)}{\mathcal{Z}^{\rm cs}(\by)}
\prod_{l=1}^LP_0(\bx_l),
\end{align*}
where $\by$ depends on the \emph{quenched} random variables $\bm{\phi},\bs, \bz$ through \eqref{eq:CSmodel}. The denominator
$\mathcal{Z}^{\rm cs}(\by)$ is the normalization, or \emph{partition function}, given by the integral of the numerator over all $\bx$ components. The {\it Gibbs averages} with respect to (w.r.t) this posterior are 
denoted by $\langle\! -\!\rangle$. For example the usual MMSE estimator is simply $\mathbb{E}[ \bX\vert \by] \!=\! \langle \bX \rangle$.  
The MI (per section) is then
\begin{align} \label{eq:true_mutual_info}
i^{\rm cs}\! \defeq\! \frac{1}{L}\mathbb{E}\Big[\!\ln\!\Big( \frac {P^{\rm cs}(\bS,\bY)}{P_0(\bS)P^{\rm cs}(\bY)}\Big)\!\Big] \!=\! -\frac{\alpha B}2\! - \!\frac{\mathbb{E}[\ln\!\mathcal{Z}^{\rm cs}(\bY)]}{L},
\end{align}
%
%
where $\mathbb{E}$ is the expectation w.r.t all the quenched random variables, $P^{\rm cs}(\by)\! =\! \mathcal{Z}^{\rm cs}(\by)(2\pi\Delta)^{-M/2}$ and $P^{\rm cs}(\bs,\by)$ is the joint distribution of the signal and the measurement. Note that $- \mathbb{E}[\ln\mathcal{Z}^{\rm cs}(\bY)]/L$ is refered as the \emph{free energy} in the statistical physics literature. 

The MMSE per section is ${\rm mmse} \!\defeq \! \mathbb{E}[\|\bS\! -\! \langle \bX\rangle\|_2^2]/L$. Unfortunately, 
this quantity is rather difficult to access directly from the 
MI. For this reason, it is more convenient to consider 
the {\it measurement MMSE} defined as ${\rm ymmse} \!\defeq \!\mathbb{E}[\|\bm{\Phi}(\bS \!-\! \langle \bX\rangle)\|_2^2]/M$ which is related to the MI by the following I-MMSE relation~\cite{shamai2005} 
\begin{align}\label{y-immse}
\frac{di^{\rm cs}}{d\Delta^{-1}} = \frac{\alpha B}{2}{\rm ymmse}.
\end{align}
Thus if we can compute the MI, we can compute the measurement MMSE and conversely. The measurement and usual MMSE's are formally related by
\begin{align}\label{xymmse}
{\rm ymmse} = \frac{{\rm mmse}}{1+ {\rm mmse}/\Delta} + \smallO_L(1),
\end{align} 
where $\lim_{L\to\infty}\smallO_L(1)\!=\!0$. As we will see we can prove and use a slightly weaker form of such a relation for a ``perturbed'' model defined in sec.~\ref{sec:partII}.

The replica method yields the \emph{replica symmetric (RS) formula} for the MI of model \eqref{eq:CSmodel}. Let $v\!\defeq\! \mathbb{E}[\bS^\intercal \bS]/L \!=\! \sum_i p_i \|\tbf{a}_i\|_2^2$, $\psi(E;\Delta)\!\defeq\! \alpha B[\ln(1\!+\! E/\Delta)\!- \!E/(E\!+\!\Delta)]/2$. The RS formula is $\lim_{L\to \infty} i^{\rm cs} \!=\! \min_{E\in [0, v]} i^{\rm RS}(E; \Delta)$ where
%
%
\begin{align}
i^{\rm RS}(E;\Delta) \defeq \psi(E;\Delta) + i(\widetilde\bS;\widetilde\bS + \widetilde\bZ\Sigma(E;\Delta)). 
\label{eq:rs_mutual_info}
\end{align}
The second term on the r.h.s is the MI for a $B$-dimensional \emph{denoising model} $\widetilde\by\!=\!\widetilde\bs\! +\! \widetilde\bz\, \Sigma$ with  $\Sigma(E;\Delta)^{-2}\!\defeq \!\alpha B/(\Delta\!+\! E)$,
\vspace{-.22cm}
\begin{align} \label{eq:i_denoising}
i(\widetilde \bS;\widetilde \bY) \defeq -\mathbb{E}_{\widetilde\bS,\widetilde\bZ}\Big[\!\ln\!\Big(\mathbb{E}_{\widetilde\bX} \Big[e^{-\!\!\sum\limits_{i=1}^B\!\!\frac{(\widetilde X_i - (\widetilde S_i + \widetilde Z_i \Sigma) )^2}{2\Sigma^2}}\Big]\Big)\Big]-\frac{B}{2},
\end{align}
$\widetilde\bZ\!\sim\!\mathcal{N}(0,\mathbf{I}_B)$ ($\mathbf{I}_B$ the $B$-dimensional identity matrix) and $\widetilde\bS,\widetilde\bX \!\sim \!P_0$. $i^{\rm RS}(E;\Delta)$ is often called the \emph{RS potential}. In the following we set $\widetilde E \!\defeq\! \argmin_{E\in [0, v]} i^{\rm RS}(E; \Delta)$.

Most interesting models have a $P_0$ such that (s.t)~\eqref{eq:rs_mutual_info}~has at most three stationary points (see the discussion in 
sec.~\ref{sec:scenarios}). Then one may show that $i^{\rm RS}(\widetilde E;\Delta)$ has \emph{at most one non-analyticity point} denoted $\Delta_{\rm RS}$. When $i^{\rm RS}(\widetilde E;\Delta)$ is analytic over $\mathbb{R}^+$ we simply set $\Delta_{\rm RS} \!=\!\infty$. The most common non-analyticity in this context is a non-differentiability point of $i^{\rm RS}(\widetilde E;\Delta)$. By virtue 
of \eqref{y-immse} (and \eqref{xymmse}) this corresponds to a jump discontinuity of the MMSE's, and one speaks of a
\emph{first order phase transition}. Another possibility is a discontinuity in higher derivatives of the 
MI, in which case the MMSE's are continuous (but non differentiable) and one speaks of \emph{higher order phase transitions}. 

The {\it main result of this paper} is a complete proof of the RS formula for $B\!=\!1$, $P_0$ discrete and s.t the RS potential \eqref{eq:rs_mutual_info} has at most three stationary points. As a consequence, we also get the large $L$ asymptotic formula for the 
measurement MMSE ${\rm ymmse}$. For general $B$ and general $P_0$ we show that $i^{\rm RS}(\widetilde E; \Delta)$ is an upper bound to $\lim_{L\to \infty}i^{\rm cs}$ (in the process we also prove the existence of the limit). We believe that with more work our method can be extended to prove the equality for this more general case. 
%
%
%
%
%
%
\subsection{Relation to previous works}
Plenty of papers about structured linear problems make use of the replica formula. In statistical physics, these date back to the late 80's with the study of the perceptron and neural networks~\cite{gardner1988space,gardner1988optimal,mezard1989space}. Of particular influence has been the work of Tanaka on CDMA~\cite{tanaka2002statistical} which has opened the way to a large set of contributions in information theory~\cite{guo2003multiuser,guo2005randomly}. In particular, the MI (or the free energy) in CS has been considered in a number of publications, e.g.~\cite{guo2009single,rangan2009asymptotic,kabashima2009typical,ganguli2010statistical,wu2012optimal,krzakala2012statistical,tulino2013support,krzakala2012probabilistic}.

In a very interesting line of work, the replica formula has emerged following the study of AMP. Again, the story of this algorithm is deeply rooted in statistical physics, with the work of Thouless, Anderson and Palmer~\cite{thouless1977solution} (thus the name ``TAP'' sometimes given to this approach). The earlier version, to the best of our knowledge, appeared in the late 80's in the context of the perceptron problem~\cite{mezard1989space}. For linear estimation, it was again developed initially in the context of CDMA~\cite{kabashima2003cdma}. It is, however, only after the application of this approach to CS~\cite{donoho2009message} that the method has gained its current popularity. Of particular importance has been the development of the rigorous proof of {\it state evolution} (SE) that allows to track the performance of AMP, using techniques developed by~\cite{bayati2011dynamics} 
 and~\cite{bayati2015universality}. Such techniques are deeply connected to the analysis of iterative forms of the TAP equations 
 by Bolthausen \cite{Bolthausen2014}. 
 Interestingly, the SE fixed points correspond to the extrema of the RS formula, 
 strongly hinting that AMP achieves the MMSE for many problems 
 where it reaches the global minimum.

While our proof technique uses SE, it is based on two important additional ingredients. The first is Guerra's interpolation method~\cite{guerra2005introduction}, that allows 
in particular to show that the RS formula yields an upper bound to the MI. This was already done for the CDMA problem in~\cite{koradamacris2007,KoradaMacris_CDMA}
(for binary signals) and here we extend this work to any $B$ and discrete $P_0$. The converse requires more work and the use of spatial coupling and threshold saturation, 
that follows recent analysis of capacity-achieving 
spatially coupled codes~\cite{kudekar2011threshold,hassani2010coupled,pfister2012,kumar2014}. Using SC in compressed sensing was proposed in~\cite{kudekar2010effect}, but it is only with the joint use of AMP that it was shown to be so powerful~\cite{krzakala2012statistical,krzakala2012probabilistic,donoho2013information}. Similar ideas have been proposed for CDMA~\cite{takeuchi2011improvement}, group testing~\cite{zhang2013non} and sparse superposition codes~\cite{barbier2014replica,barbier2015approximate,barbier2016proof}. 

The authors have recently applied a similar strategy to the factorization of low rank matrices~\cite{krzakala2016mutual,XXT}. This, we believe, shows that the developed techniques and results proved 
in this paper are not only relevant for random linear estimation, but also in a broader context, opening the way to prove many results on estimation problems previously obtained with the heuristic replica method.

Finally we wish to point out that we have received a private communication 
from~\cite{private} who reached at the same time similar results using a very different approach. 
%
\subsection{Approximate message-passing and state evolution}
AMP is deeply linked to \eqref{eq:rs_mutual_info}. Its asymptotic performance for the CS model can be rigorously tracked by SE in the scalar $B\!=\!1$ case~\cite{bayati2011dynamics,donoho2013information}. The vectorial $B\!\ge\!2$ case requires extending the SE analysis rigorously, which at the moment has not been done to the best of our knowledge.\footnote{We thank Marc Lelarge and Andrea Montanari for clarifications on this.} Nevertheless, we conjecture that SE (see \eqref{recursion-uncoupled-SE} below) tracks AMP for any $B$. This is numerically confirmed in~\cite{barbier2015approximate} and proven for power allocated sparse superposition codes~\cite{rush2015capacity}.

Denote $E^{(t)}\!\defeq \lim_{L\to\infty}\!\mathbb{E}[\|\bS \!- \! {\widehat{\bS}}\!\!~^{(t)}\|^2_2]/L$ the asymptotic average MSE obtained by AMP at iteration $t$, $\widehat{\bs}\!\!~^{(t)}$ being the AMP estimate at $t$. Denote the MMSE associated with the denoising model (introduced in sec \ref{sec:settings}) by ${\rm mmse}(\Sigma^{-2}) \!\defeq\! \mathbb{E}[\|\widetilde\bS\!-\! \mathbb{E}[\bX\vert \widetilde\bS \!+\! \widetilde\bZ\Sigma ]\|^2_2]$. The SE recursion tracking AMP is
\begin{align}\label{recursion-uncoupled-SE}
E^{(t+1)} = {\rm mmse}(\Sigma(E^{(t)}; \Delta)^{-2}),
\end{align}
with the initialisation $E^{(0)} \!=\! v$. Monotonicity 
properties of ${\rm mmse}(\Sigma^{-2})$ imply that $E^{(t)}$ is a decreasing sequence s.t $\lim_{t\to \infty}E^{(t)} \!=\! E^{(\infty)}$ exists.
%
%
%
%
%
Let us give a natural definition for the AMP threshold.

\begin{definition}[AMP algorithmic threshold]\label{algo-thresh-def} 
$\Delta_{\rm AMP}$ is the supremum of all $\Delta$ s.t the fixed point equation associated with \eqref{recursion-uncoupled-SE} has a \emph{unique} solution for all noise values in $[0, \Delta]$.
\end{definition}
 
\begin{remark}[SE and $i^{\rm RS}$ link] \label{remark:extrema_irs_fpSE}
The extrema of \eqref{eq:rs_mutual_info} correspond to the fixed points of the SE recursion \eqref{recursion-uncoupled-SE}.
Thus $\Delta_{\rm AMP}$ is also the smallest solution of 
$\partial i^{\rm RS}/ \partial E\!=\! \partial^2 i^{\rm RS}/\partial E^2 \!=\! 0$; in other words it is 
the ``first'' horizontal inflexion point appearing in $i^{\rm RS}(E;\Delta)$ when $\Delta$ increases. 
\end{remark}
\subsection{Results: mutual information and measurement MMSE}
\label{sec:results}
Our first result states that the minimum of \eqref{eq:rs_mutual_info} upper bounds the asymptotic MI.
\begin{thm}[Upper Bound] \label{th:upperbound}
Assume model \eqref{eq:CSmodel} with any $B$ and discrete prior $P_0$. Then 
\begin{equation*}
\lim_{L\to \infty} i^{\rm cs}  \le \min_{E\in [0, v]} i^{\rm RS} (E;\Delta).
\end{equation*}
\end{thm}

This result generalizes the one already obtained for CDMA in~\cite{koradamacris2007,KoradaMacris_CDMA}, and we note that a further generalization 
to more general priors that are mixtures of discrete and absolutely continuous parts (as long as the support is bounded) can also be achieved
without any major change in our proof. The next result yields the equality in the scalar case.
\begin{thm}[One letter formula for $i^{\rm cs}$] \label{th:replica}
Take $B=1$ and assume $P_0$ is a discrete
prior s.t $i^{\rm RS}(E;\Delta)$ in \eqref{eq:rs_mutual_info} has at most three stationary points (as a function of $E$). 
Then for any $\Delta$ the RS formula is true, that is
\begin{align}\label{eq:replicaformula}
\lim_{L\to \infty} i^{\rm cs}  = \min_{E\in [0, v]} i^{\rm RS}(E;\Delta).\end{align}
\end{thm}
\vspace{.2cm}
It is conceptually useful to define the following threshold.
\begin{definition}[Information theoretic threshold]\label{def-delta-opt} 
Define $\Delta_{\rm Opt} \!\defeq\!\sup\{\Delta \ \text{s.t}\ \lim_{L\to \infty}i^{\rm cs} \ \text{is analytic in} \ ]0, \Delta[\}$.
\end{definition}

Theorem~\ref{th:replica} gives us an explicit formula to {\it compute} the information theoretical threshold $\Delta_{\rm Opt} \!=\! \Delta_{\rm RS}$.

Using \eqref{y-immse} and the theorem~\ref{th:replica}, we obtain the following.
\begin{corollary}[measurement MMSE]\label{cor:MMSE}
Under the same assumptions as in theorem~\ref{th:replica} and for any
$\Delta\!\neq\! \Delta_{\rm RS}$  the measurement MMSE for model \eqref{eq:CSmodel} satisfies 
\begin{align*}
\lim_{L\to \infty}{\rm ymmse}=\frac{\widetilde E}{1+\widetilde E/\Delta},
\end{align*}
where $\widetilde E$ is the unique global minimum of $i^{\rm RS}(E;\Delta)$.
\end{corollary}

The proofs of theorems~\ref{th:upperbound} and \ref{th:replica} are discussed in sec.~\ref{sec:partII} and \ref{sec:partIII}. We conjecture that theorem~\ref{th:replica} and corollary~\ref{cor:MMSE} hold for any $B$. Their proofs require a control of AMP by SE, a result that (to our knowledge) is currently available in the literature only for $B\!=\!1$. Proving SE for all $B$ would imply these results for the vectorial case, and we believe that this is not out of reach.

Two important and related issues that can be tackled with our methods are the following. 
Firstly, corollary~\ref{cor:MMSE} should extend to the usual MMSE instead of the measurement one (i.e., $\lim_{L\to \infty}{\rm mmse}\! =\! {\widetilde E}$). 
Secondly, as $L\!\to\!\infty$, AMP initialized without any 
knowledge other than $P_0$ yields upon convergence the asymptotic measurement \emph{and} usual MMSE if $\Delta \!<\!\Delta_{\rm AMP}$ or $\Delta\! >\! \Delta_{\rm RS}$. These problems will not be discussed further here due to lack of space and we will come back to them in a future contribution.  Another direction for generalization is to consider $P_0$ a mixture of discrete and absolutely continuous parts.
\subsection{The single first order phase transition scenario} \label{sec:scenarios}
In this contribution, we assume that $P_0$ is discrete and s.t \eqref{eq:rs_mutual_info} has \emph{at most} three stationnary points. 
Let us briefly discuss what this hypothesis entails. 

Three scenarios are possible:  $\Delta_{\rm AMP}\! <\!\Delta_{\rm RS}$ (one \emph{first order} phase transition); 
$\Delta_{\rm AMP}\!=\!\Delta_{\rm RS}\! <\!\infty$ (one \emph{higher order} phase transition); 
$\Delta_{\rm AMP}\!=\!\Delta_{\rm RS}\!=\! \infty$ (no phase transition). In the sequel we will consider the 
most interesting (and challenging) first order phase transition case where a gap between the algorithmic AMP and 
information theoretic performance appears.
The cases of no or higher order phase transition, which present no algorithmic gap, follow as special cases from
our proof. It should be noted that in these two cases spatial coupling is not really needed and the proof 
can be achieved by an ``area theorem'' as already showed in \cite{MontanariTse06}.


Recall the notation $\widetilde E(\Delta) \!=\! \argmin_{E\in[0,v]} i^{\rm RS}(E;\Delta)$. At $\Delta_{\rm RS}$, when the $\argmin$ is a set with two elements, one can think of it as a discontinuous function.

The picture for the stationary points of \eqref{eq:rs_mutual_info} is as follows. 
For $\Delta \!<\!\Delta_{\rm AMP}$ there is a unique stationary point 
which is a global minimum $\widetilde E$ and we have $\widetilde E \!=\! E^{(\infty)}$. At 
$\Delta_{\rm AMP}$ $i^{\rm RS}$ develops
a horizontal inflexion point, and for $\Delta_{\rm AMP}\!<\!\Delta\! <\! \Delta_{\rm RS}$ there are three stationary points: a local 
minimum corresponding to $E^{(\infty)}$, a local maximum, and the global minimum $\widetilde E$. It is not 
difficult to argue that 
$\widetilde E \!<\! E^{(\infty)}$ in the interval $\Delta_{\rm AMP}\!<\!\Delta \!<\! \Delta_{\rm RS}$. At $\Delta_{\rm RS}$ 
the local and global minima switch roles, so at this point the global minimum
$\widetilde E$ has a jump discontinuity. For all $\Delta\!>\!\Delta_{\rm RS}$ there is at least one stationary point which 
is the global minimum $\widetilde E$ and $\widetilde E\!=\! E^{(\infty)}$ (the other stationary points can merge and annihilate each other 
as $\Delta$ increases). 

Finally we note that with the help of the implicit function theorem for real analytic functions we can show that $\widetilde E(\Delta)$ is an analytic function of $\Delta$ except at $\Delta_{\rm RS}$. Therefore $i^{\rm RS}(\widetilde E, \Delta)$ is analytic in $\Delta$ except at $\Delta_{\rm RS}$.

\section{Proof strategy}
\label{sec:partII}
\subsection{A general interpolation} \label{sec:generalInterpolation}
We have already seen that the RS potential \eqref{eq:rs_mutual_info} involves the MI of a denoising model (see above \eqref{eq:i_denoising}). One of the main tools that we use is an interpolation between a simple denoising model and the original CS model \eqref{eq:CSmodel}.
Consider a set of observations $[\by, \widetilde\by]$ from the following channels (here $\bZ\!\sim\!\mathcal{N}(0,\mathbf{I}_M),\widetilde \bZ\!\sim\!\mathcal{N}(0,\mathbf{I}_N)$)
\begin{align*} 
\begin{cases}
\by = \bm{\phi}\bs + \bz\frac{1}{\sqrt{\gamma(t)}}, \\ 
\widetilde{\by} = \bs + \widetilde{\bz}\frac{1}{\sqrt{\lambda(t)}} ,
\end{cases}
\end{align*}
where $t\!\in\! [0, 1]$ is the interpolating parameter and the signal-to-noise (snr) functions $\gamma(t)$ and $\lambda(t)$ (let us call these snr despite the signal power $v$ may be $\neq\!1$) satisfy the constraint
\begin{align} 
 \frac{\alpha B}{\gamma(t)^{-1} + E} + \lambda(t) = \frac{\alpha B}{\Delta +E} = \Sigma(E;\Delta)^{-2},
 \label{snrconstraint}
\end{align}
and $\gamma(0) \!=\!  \lambda(1) \!=\! 0$, $\gamma(1) \!=\! 1/\Delta$, $ \lambda(0) \!=\! \Sigma(E;\Delta)^{-2}$. We also require $\gamma(t)$ to be strictly increasing and $\lambda(t)$ strictly decreasing.

In order to prove concentration properties that are needed in our proofs, we will actually work with a more complicated {\it perturbed interpolated model} where we add a set of extra observations 
that come from another ``side channel'' denoising model $\widehat \by \!=\! \bs \!+\! \widehat \bz/\sqrt h$, $\widehat\bZ\!\sim\!\mathcal{N}(0,\mathbf{I}_N)$.
Here the snr $h$ is ``small'' and one should keep in mind that it will be removed in the process of the proof, i.e., $h\!\to\! 0$ (from above). 

Define $\mathring{\by}\!\defeq\! [\by, \widetilde \by,\widehat \by]$ as the concatanation of all observations. Our central object of study is 
the posterior of this general perturbed interpolated model
\begin{align}
P_{t,h}(\bx|\mathring \by) = \frac{\exp\Big(-\mathcal{H}_{t,h}(\bx|\mathring{\by})\Big)}{\mathcal{Z}_{t,h}(\mathring{\by})}
\prod_{l=1}^L P_0(\bx_l),
\label{gibbs-general}
\end{align}
where 
the \emph{Hamiltonian} is
\begin{align} 
&\mathcal{H}_{t,h}(\bx|\mathring{\by}) \defeq \frac{h}{2}\sum_{i=1}^N \!\Big(\bar x_i \!-\! \frac{\widehat z_i}{\sqrt{h}} \Big)^2 \label{eq:int_hamiltonian}\\
&+ \frac{\gamma(t)}{2}\sum_{\mu=1}^M\! \Big([\bm{\phi}\bar \bx]_\mu\!\!-\! 
\frac{z_{\mu}}{\sqrt{\gamma(t)}}\Big)^2\!\! +\! \frac{\lambda(t)}{2}\sum_{i=1}^N \!\Big(\bar x_i\!-\!\frac{\widetilde z_i}{\sqrt{\lambda(t)}}\Big)^2, \nonumber
\end{align}
and $\mathcal{Z}_{t,h}(\mathring{\by})$ is the partition function (the integral of the numerator over all $\bx$ components). Note that
the quenched random variables $\bm{\phi}$, $\bz$, $\widetilde{\bz}$ and $\widehat{\bz}$ are all independent. As before, expectations w.r.t the Gibbs measure \eqref{gibbs-general} are denoted $\langle \!-\!\rangle_{t,h}$, expectations w.r.t the quenched random variables by $\mathbb{E}$. 

The MI $i_{t,h}$ for the perturbed interpolated model is defined 
similarly as \eqref{eq:true_mutual_info}.
Note that $i_{1,0}\!=\!i^{\rm cs}$.
\begin{remark}[snr conservation] \label{remark:snrcons}
Constraint \eqref{snrconstraint}, or \emph{snr conservation}, is essential. It expresses that as $t$ decreases from $1$ to $0$, we slowly decrease the snr of the CS measurements and 
make up for it in the denoising model. When $t\!=\!0$ the snr vanishes for the CS model, and no information is available about $\bs$ from the compressed measurements, information comes only from the denoising model. Instead at $t\!=\!1$ the noise is infinite in the denoising model and letting also $h\!\to\!0$ we recover the CS model. 
 
This constraint can be interpreted as follows. Given a CS model of snr $\Delta^{-1}$, by remark~\ref{remark:extrema_irs_fpSE} and \eqref{recursion-uncoupled-SE}, the global minimum of \eqref{eq:rs_mutual_info} is the MMSE of an ``effective'' denoising model of snr $\Sigma(E;\Delta)^{-2}$. Therefore, 
the interpolated model \eqref{eq:int_hamiltonian} (at $h\!=\!0$) is asymptotically equivalent (in the sense that it has the same MMSE)
as two independent denoising models: an ``effective'' one of snr $\Sigma(E;\gamma(t)^{-1})^{-2}$ associated with the CS model, 
and another one with snr $\lambda(t)$. Proving theorem~\ref{th:replica} requires 
the interpolated model to be designed s.t its MMSE equals the MMSE of the CS model \eqref{eq:CSmodel} for almost all $t$. Knowing that the estimation of $\bs$ in the interpolated model comes 
from \emph{independent} channels, this MMSE constraint induces \eqref{snrconstraint}.
\end{remark}
\begin{remark}[Nishimori identity]\label{remark:Nishi}
We place ourselves in the Bayes optimal 
setting where $P_0,\Delta, \gamma(t), \lambda(t)$ and $h$ are known. The perturbed interpolated model is carefully designed, that is each of the three terms in \eqref{eq:int_hamiltonian} corresponds to a ``physical'' channel model, s.t the \emph{Nishimori identity} holds. This remarquable and general identity (from which many convenient ``sub-identities'' follow) plays an important role in our calculations.
For any function $g(\bx, \bs)$: if $\bs$ is the signal, then
\begin{align*}
&\mathbb{E}[\langle g(\bX, \bS)\rangle_{t,h}] = \mathbb{E}[\langle g(\bX, \bX')\rangle_{t,h}], 
\end{align*}
where $\bX, \bX'$ are i.i.d vectors distributed according to the product measure of \eqref{gibbs-general}. We abuse notation here by 
denoting the posterior measure for $\bX$ and the product measure for $\bX, \bX^\prime$ with the 
same bracket $\langle\! -\! \rangle_{t, h}$. 
\end{remark}
\subsection{Various MMSE's}
We will need the following I-MMSE lemma that straightforwardly extends to the perturbed interpolated model the usual I-MMSE theorem~\cite{GuoVerduShamai_IMMSE} for the vectorial denoising model. Let ${\rm ymmse}_{t,h} \!\defeq \!\mathbb{E}[\|\bm{\Phi}(\bS\!-\!\langle \bX\rangle_{t,h})\|_2^2]/M$. Then

\begin{lemma}[I-MMSE] \label{lemma:Immse}
$di_{1,h}/d\Delta^{-1} \!=\! {\rm ymmse}_{1,h}\alpha B/2$.
\end{lemma}

Let us give a useful link between ${\rm ymmse}_{t,h}$ and the usual MMSE
$E_{t,h} \!\defeq \!\mathbb{E}[\|\bS\!-\! \langle\bX\rangle_{t,h}\|_2^2]/L$. 
For the perturbed interpolated model ($\bm{\phi}$ i.i.d Gaussian), the following holds (proof sketch in sec.~\ref{sec:ymmse}).

\begin{lemma}[MMSE relation] \label{lemma:MSEequivalence}
For almost every (a.e) $h$, ${\rm ymmse}_{t,h} \!=\!E_{t,h}(1 + \gamma(t)E_{t,h})^{-1}\!+\!\smallO_L(1)$.
\end{lemma}

In this lemma $\lim_{L\to \infty}\smallO_L(1)\!=\!0$. However in our proof $\smallO_L(1)$ is not uniform in $h$ and diverges as $h^{-1/2}$ as $h\!\to\! 0$. For this reason we cannot interchange the limits $L\!\to\!\infty$ and $h\!\to\!0$. This is not only a technicality, because in the presence of a first order phase transition one has to somehow deal with the discontinuity at $\Delta_{\rm RS}$. 
\subsection{The integration argument} \label{sec:integrationArgument}
We first remark that 
AMP is sub-optimal. Thus when used for inference over the CS model \eqref{eq:CSmodel} (with $t\!=\!1, h\!=\!0)$
one gets $\limsup_{L\to\infty}E_{1,0} \leq E^{(\infty)}$. 
Adding new measurements 
can only improve optimal inference thus $E_{1, h}\leq E_{1,0}$ and 
$\limsup_{L\to\infty}E_{1,h} \leq E^{(\infty)}$.
Combining this with lemma~\ref{lemma:MSEequivalence} and using that $E(1
\!+\!E/\Delta)^{-1}$ is an increasing function of $E$, one gets that for a.e $h$
\begin{align}
 \limsup_{L\to\infty} \frac{\alpha B}{2}{\rm ymmse}_{1,h} \leq \frac{\alpha B}{2}\frac{E^{(\infty)}}{1+ E^{(\infty)}/\Delta}.
 \label{ymmseinequality}
\end{align}

Now let us look at the case $\Delta\! < \! \Delta_{\rm AMP}$ first. In this noise regime $E^{(\infty)}\!=\!\widetilde{E}$ the global minimum 
of $i^{\rm RS}$ (see remark~\ref{remark:extrema_irs_fpSE}) so we replace $E^{(\infty)}$ by $\widetilde E$ in the r.h.s of \eqref{ymmseinequality}. Furthermore by a rather explicit differentiation one checks that 
$di^{\rm RS}(\widetilde E;\Delta)/d\Delta^{-1} \!=\! (\alpha B/2)\widetilde E (1\!+\! \widetilde E/\Delta)^{-1}$. Then, using also lemma~\ref{lemma:Immse} the inequality \eqref{ymmseinequality} becomes
\begin{align*}
\limsup_{L\to\infty}\! \frac{di_{1,h}}{d\Delta^{-1}} 
\!\leq \!
\frac{di^{\rm RS}(\widetilde E;\Delta)}{d\Delta^{-1}} 
{\rm ~or~}
\frac{di^{\rm RS}(\widetilde E;\Delta)}{d\Delta} \!\leq\! \liminf_{L\to\infty}\! \frac{di_{1,h}}{d\Delta}.
\end{align*}
Integrating the last inequality over $[0, \Delta]\subset [0, \Delta_{\rm AMP}]$ and using Fatou's lemma we get 
\begin{align}
i^{\rm RS}(\widetilde E;\Delta)\! -\!i^{\rm RS}(\widetilde E;0) \leq 
\liminf_{L\to\infty} (i_{1,h}|_{\Delta} \!-\!  i_{1,h}|_{0}). \label{eq:beforeSimplifying}
\end{align}
It is easy to see that $i_{t,h}$ is concave, and thus continuous, in $h$. Our interpolation proofs show superadditivity of this sequence so that by Fekete's lemma the limit $L\!\to\!\infty$ of $i_{t,h}$ exists, and is thus also concave and continuous in $h$. As a consequence we can take the limit $h\!\to\! 0$ in the last inequality and permute the limits
$h\!\to\! 0$ and $L\!\to \!\infty$. 
Furthermore, for discrete priors, one can show that $i^{\rm RS}(\widetilde E;0) \!=\! \lim_{L\to\infty} i^{\rm cs}|_{\Delta=0} \!=\! H(\bS)$ the 
Shannon entropy of $\bS\!\sim\! P_0$. So we obtain from \eqref{eq:beforeSimplifying} that $i^{\rm RS}(\widetilde E;\Delta) \!\le \!\lim_{L\to\infty} i^{\rm cs}|_{\Delta}$, 
which 
combined with theorem~\ref{th:upperbound}, yields theorem~\ref{th:replica} for all $\Delta\! \in\! [0,\Delta_{\rm AMP}]$. 

Notice that $\Delta_{\rm AMP} \!\leq\! \Delta_{\rm Opt}$. While this might seem clear, 
it follows from $\Delta_{\rm RS} \!\geq \!\Delta_{\rm AMP}$ (by their definitions) which together with 
 $\Delta_{\rm AMP} \!>\! \Delta_{\rm Opt}$ would imply from theorem~\ref{th:replica} that $\lim_{L\to \infty} i^{\rm cs}$ is analytic at $\Delta_{\rm Opt}$, a contradiction.

\emph{Assume for a moment that} $\Delta_{\rm Opt} \!=\! \Delta_{\rm RS}$. Thus both $\lim_{L\to\infty}i^{\rm cs}$ and $i^{\rm RS}(\widetilde E; \Delta)$ are analytic until $\Delta_{\rm RS}$
which, since they are equal on $[0, \Delta_{\rm AMP}]\!\subset\! [0, \Delta_{\rm RS}]$,
implies by unicity of the analytic continuation that they are equal for all $\Delta\!<\!\Delta_{\rm RS}$. 
Concavity in $\Delta$ implies continuity of $\lim_{L\to \infty}i^{\rm cs}$ which allows to conclude that theorem 
\ref{th:replica} holds at $\Delta_{\rm RS}$ too.

Now consider $\Delta\! \ge\! \Delta_{\rm RS}$. 
Then again $E^{(\infty)}\!=\!\widetilde E$ the global 
minimum of $i^{\rm RS}$. We can start again from \eqref{ymmseinequality} with $E^{(\infty)}$ replaced by $\widetilde E$
and apply a similar integration argument with the integral now running from $\Delta_{\rm RS}$ to $\Delta$. The validity of the replica formula at $\Delta_{\rm RS}$ that we just proved above is crucial to complete this argument.

It remains to show $\Delta_{\rm Opt} \!=\! \Delta_{\rm RS}$. This is where SC and threshold saturation come as new crucial ingredients. 
\subsection{Proof of $\Delta_{\rm Opt}\!=\!\Delta_{\rm RS}$ using spatial coupling} \label{sec:deltaopt_deltars}
\begin{figure}[!t]
\centering
\includegraphics[width=.235\textwidth]{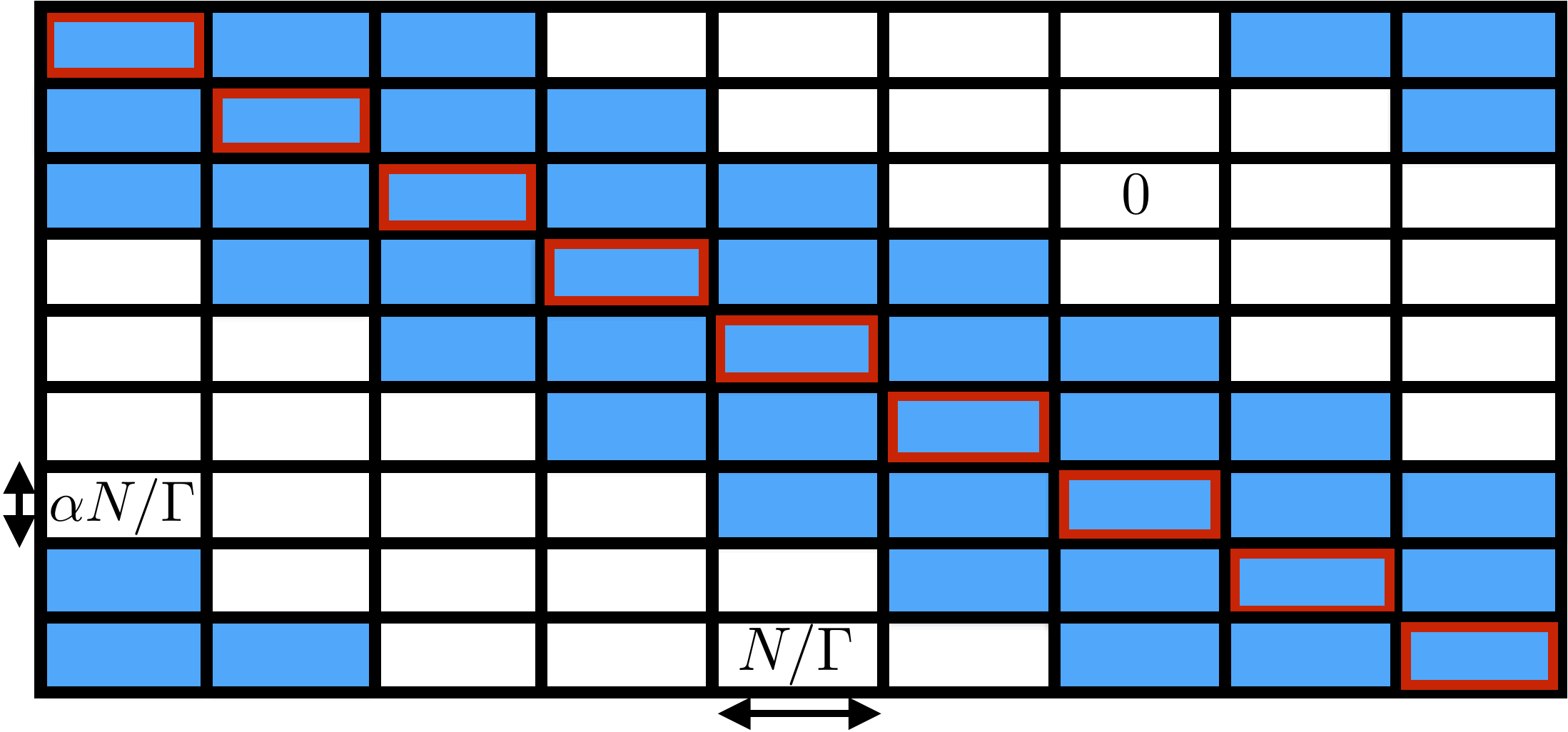}
\includegraphics[width=.236\textwidth]{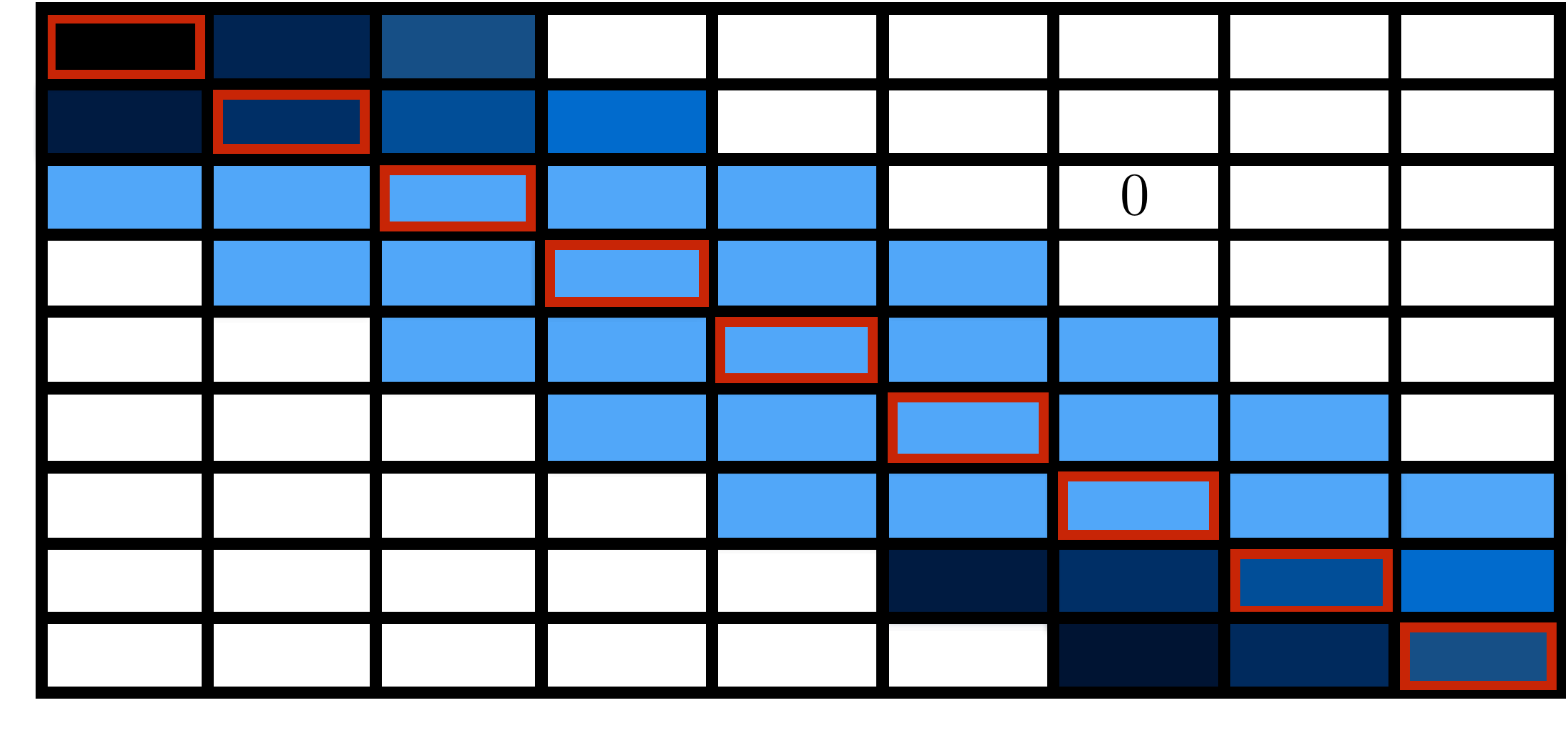}
\caption{Spatially coupled measurement matrices $\in\!\mathbb{R}^{M\times N}$ with a ``band diagonal'' structure. They are made of $\Gamma\!\times\! \Gamma$ \emph{blocks} indexed by $(r,c)$ (here $\Gamma\!=\!9$), each with $N/\Gamma$ columns and $M/\Gamma$ rows. The i.i.d entries inside block $(r,c)$ are $\mathcal{N}(0,J_{r,c}/L)$. The \emph{coupling strength} is controled by the variance matrix $\tbf{J}$. We consider two slightly different constructions. The \tbf{periodic matrix} (left): it has $w$ forward and $w$ backward \emph{coupling blocks} (here $w\!=\!2$) with $J_{r,c}\!=\!\Gamma/(2w\!+\!1)$ if $|r\!-\!c|\!\le\!w$ $(\!\!\!\!\mod \Gamma$), $0$ else (white blocks with only zeros). The \tbf{opened matrix} (right): the \emph{coupling window} $w$ remains unchanged except at the boundaries where the periodicity is broken. Moreover the coupling strength is stronger at the boundaries (darker color).}
\label{fig:opSpCoupling}
\end{figure}
\underline{\smash{\tbf{Spatial coupling:}}}
In order to show the equality of the thresholds, we need the introduction 
of two closely related \emph{spatially coupled CS models}. Their construction is described by fig.~\ref{fig:opSpCoupling}
which shows two measurement matrices replacing the one of the CS model \eqref{eq:CSmodel}, whose structure induce a natural \emph{block} decomposition of the signal $\bs$ in $\Gamma$ blocks made of $L/\Gamma$ sections. 
On the left the matrix corresponds to taking periodic boundary conditions. This is called the 
{\it periodic} SC model. On the right the SC matrix is opened. This corresponds to the {\it seeded} SC model because for this model, we assume that the signal components 
are known at its boundary blocks $\in\!\mathcal{B}$, which size $|{\cal B}|$ is of order $w$ blocks (see~\cite{barbier2016proof,barbier2016threshold} for precise statements). 
The stronger variance at the boundaries of opened matrices help this \emph{information seed} to trigger a reconstruction wave that propagates inward the signal. This phenomenon is what allows SC to reach such good results, namely reconstruction by AMP at low $\alpha$. 

\underline{\smash{\tbf{Threshold saturation:}}}
AMP performance, when SC matrices are used, is tracked by an MSE \emph{profile} $\tbf{E}^{(t)}$: a 
vector $\in\![0,v]^\Gamma$ whose components are MSE's describing the quality of the reconstructed signal, see \cite{barbier2016proof,barbier2016threshold} for details. 

Consider the seeded SC model. The MSE profile $\tbf{E}^{(t)}$ can be asymptotically computed by SE. The precense of the seed is reflected by $E_r^{(t)} \!= \!0$ for all $t$ if $r\!\in\!\mathcal{B}$, else
\begin{align}
&E_r^{(t+1)} \!=\! \frac{1}{\Gamma}\sum_{c=1}^\Gamma \!J_{r,c} \, {\rm mmse}(\Sigma_c(\tbf{E}^{(t)}; \Delta)^{-2})\ \text{if} \ r\notin \mathcal{B}, \label{eq:SEc} \\ 
&\Sigma_c(\tbf{E}; \Delta)^{-2} \defeq \frac{\alpha B}{\Gamma}\sum_{r=1}^{\Gamma} \frac{J_{r,c}}{\Delta +E_r},\label{eq:SEcoupled2}
\end{align}
with initialization $E_r^{(0)} \!=\! v \ \forall \ r\!\in\!\{1,\ldots,\Gamma\}\backslash \mathcal{B}$, as required by 
AMP. Denote $\tbf{E}^{(\infty)}$ the fixed point 
of this SE recursion \eqref{eq:SEc}, \eqref{eq:SEcoupled2} and $E_{\rm good}(\Delta)$ the smallest solution of the fixed point equation associated with the SE recursion \eqref{recursion-uncoupled-SE}. The \emph{algorithmic threshold of the seeded SC model} is $\Delta_{\rm AMP}^{\rm c}\!\defeq\! \liminf_{w\to\infty}\liminf_{\Gamma\to\infty}
\sup\{\Delta\!>\!0 \ \text{s.t}\ E_r^{(\infty)}\!\le\!E_{\rm good}(\Delta)\, \forall\, r\}$ where ${\liminf}$ is taken 
along sequences where {\it first} $\Gamma\! \to \!\infty$ and {\it then} $w\!\to\!\infty$. It is proved in \cite{barbier2016proof,barbier2016threshold} by three of us that when AMP is used for inference on seeded SC models, \emph{threshold saturation} occurs, that is:
\begin{lemma}[Threshold saturation]\label{lemma:threshSat}
$\Delta_{\rm AMP}^{\rm c}\geq \Delta_{\rm RS}$.
\end{lemma}
Note that in fact the equality holds, but we shall not need it. 

\underline{\smash{\tbf{Invariance of the optimal threshold:}}}
Call the MI per section for the periodic and seeded SC models, respectively, $i^{\rm per}_{\Gamma,w}$ and $i_{\Gamma, w}^{\rm seed}$. Using an interpolation we can 
show the following asymptotic equivalence property, see sec.~\ref{sec:mutualInfoSCmodel}.
\begin{lemma}[Invariance of the MI] \label{lemma:modelsEquiv}
The following limiting mutual informations exist and are equal 
for any $\Gamma$, $w\!\in\!\{0,\ldots,(\Gamma\!-\!1)/2\}$: $\lim_{L\to \infty} i_{\Gamma, w}^{\rm per}\! = \!\lim_{L\to \infty} i^{\rm cs}$. Moreover for any fixed $w$, $\lim_{\Gamma\to \infty}\lim_{L\to \infty} i_{\Gamma, w}^{\rm seed}\! = \!\lim_{L\to \infty} i^{\rm cs}$ too. 
\end{lemma}

This implies straightforwardly that the optimal 
threshold $\Delta_{\rm Opt}^{\rm c}$ of the seeded SC model, defined as the first non-analyticity point (as $\Delta$ increases) of its asymptotic MI (with the appropriate order of limits), is the same as the one of the CS model, namely $\Delta_{\rm Opt}^{\rm c}\!=\! \Delta_{\rm Opt}$.

\underline{\smash{\tbf{The inequality chain:}}}
We claim the following
\begin{align*}
\Delta_{\rm RS} \le \Delta_{\rm AMP}^{\rm c} \le \Delta_{\rm Opt}^{\rm c} = \Delta_{\rm Opt} \le \Delta_{\rm RS},
\end{align*}
and therefore $\Delta_{\rm Opt} = \Delta_{\rm RS}$. 

The first inequality is lemma~\ref{lemma:threshSat}. The second follows from sub-optimality of AMP for the seeded SC model. The equality follows from lemma~\ref{lemma:modelsEquiv} (together with 
the discussion below it). The last inequality requires a final argument that we now explain.  

Recall that $\Delta_{\rm Opt}\! <\! \Delta_{\rm AMP}$ is not possible. Let us show that $\Delta_{\rm RS}\!\in \,]\Delta_{\rm AMP}, \Delta_{\rm Opt}[$ is also impossible. We proceed by contradiction so we suppose this is true. Then each side of \eqref{eq:replicaformula} are analytic on $]0, \Delta_{\rm RS}[$ and 
since they are equal for $]0, \Delta_{\rm AMP}[ \subset ]0, \Delta_{\rm RS}[$, they must 
be equal on the whole range $]0, \Delta_{\rm RS}[$ and also at $\Delta_{\rm RS}$ by continuity. 
For $\Delta\!> \!\Delta_{\rm RS}$ the fixed point of SE 
is $E^{(\infty)} \!=\! \widetilde{E}$ the global minimum of $i^{\rm RS}(E;\Delta)$, hence, the integration argument can be used once more on an interval $[\Delta_{\rm RS}, \Delta]$ which implies that \eqref{eq:replicaformula} holds for all $\Delta$. But then $i^{\rm RS}(\widetilde E;\Delta)$ is analytic at $\Delta_{\rm RS}\!\in \,]\Delta_{\rm AMP}, \Delta_{\rm Opt}[$ which is a \emph{contradiction}. 

\section{Proofs}
\label{sec:partIII}
\subsection{Upper bound using Guerra's interpolation method}\label{sec:upperBound}
The goal of this section is to sketch the proof of theorem~\ref{th:upperbound}. 
First note that the denoising model has been designed specifically so that
\begin{align}
i_{0,0} = i^{\rm RS}(E;\Delta) - \psi(E;\Delta), \label{eq:relation_fden_frs}
\end{align}
see \eqref{eq:rs_mutual_info}.
By the fundamental theorem of calculus, we have
$
i_{1,h} \!=\! i_{0,h} + \int_0^1 dt (di_{t,h}/dt).
$
Using \eqref{eq:relation_fden_frs} and a bit of algebra this is equivalent to 
\begin{align}
i_{1,h} &= i^{\rm RS}(E;\Delta)+ (i_{0,h} -i_{0,0}) + \int_0^1 dt R_{t, h} \, ,
\label{eq:fcs_fdedn_plus_remainder}\\
R_{t, h} &= \frac{di_{t,h}}{dt} - \frac{\alpha B}{2} \frac{d\gamma(t)}{dt} \frac{\gamma(t)E^2}{(1 + \gamma(t)E)^2}.
\label{eq:remainder}
\end{align}
We derive a useful expression for the remainder $R_{t,h}$ which shows that it is negative up to a negligible term. 
Straightforward differentiation gives 
$d i_{t,h}/dt \!= \!({\cal A} + {\cal B})/(2L)$ where
\begin{align}
{\cal A} &= \frac{d\gamma(t)}{dt}\sum_{\mu=1}^M \mathbb{E}\Big[\Big\langle [\bm{\Phi}\bar \bX]_\mu^2 - \gamma(t)^{-\frac{1}{2}}[\bm{\Phi}\bar \bX]_\mu Z_{\mu} \Big\rangle_{t,h}\Big], \label{eq:A_twoterms}\\
{\cal B} &= \frac{d\lambda(t)}{dt} \sum_{i=1}^N \mathbb{E}\Big[\Big\langle \bar X_i^2 - \lambda(t)^{-\frac{1}{2}}\bar X_i \widetilde Z_i\Big\rangle_{t,h}\Big]. \nonumber
\end{align}
These two quantities can be simplified using 
Gaussian integration by parts. For example, integrating by parts w.r.t $Z_\mu$,
\begin{align}
\gamma(t)^{-\frac{1}{2}}\mathbb{E}_{\bZ}[ \langle [\bm{\phi}\bar \bX]_\mu \rangle_{t,h}Z_\mu] &= \mathbb{E}_{\bZ}[\langle [\bm{\phi}\bar \bX]_\mu^2 \rangle_{t,h} - \langle [\bm{\phi}\bar \bX]_\mu \rangle_{t,h}^2], \nonumber
\end{align}
which allows to simplify $\mathcal{A}$. For $\mathcal{B}$ we proceed similarly with an integration by parts w.r.t $\widetilde Z_i$, and find
\begin{align*}
{\cal A} = \frac{d\gamma(t)}{dt} \sum_{\mu=1}^M \mathbb{E}[\langle [\bm{\Phi}\bar \bX]_\mu \rangle_{t,h}^2] , \quad {\cal B} =\frac{d\lambda(t)}{dt} \sum_{i=1}^N \mathbb{E}[\langle \bar X_i \rangle_{t,h}^2 ].
\end{align*}
Now, recalling the definitions of $E_{t,h}$ and ${\rm ymmse}_{t,h}$, using lemma~\ref{lemma:MSEequivalence} and the snr conservation relation \eqref{snrconstraint} we see that these two formulas are equivalent, respectively, to 
\begin{align}
\frac{{\cal A}}{2L} \!&=\! \frac{\alpha B}{2} \frac{d\gamma}{dt} {\rm ymmse}_{t,h} \!= \!\frac{d\gamma}{dt}\frac{\alpha B E_{t,h}}{2(1\!+\!\gamma E_{t,h})} \!+ \!\smallO_L(1), \label{eq:secondTerm_fint_A}\\
\frac{\mathcal{B}}{2L} \!&=\! \frac{d\lambda}{dt} \frac{E_{t,h}}{2} \!=\! - \frac{d\gamma}{dt}\frac{\alpha B E_{t,h}}{2(1\!+\!\gamma E)^{2}}. \label{eq:secondTerm_fint}
\end{align}
%
%
%
Notice that the first relation is true for a.e $h$. Finally, combining \eqref{eq:remainder}-\eqref{eq:secondTerm_fint} gives
for a.e $h$
\begin{align} \label{eq:R_negative}
R_{t,h} \!=\! - \frac{d\gamma(t)}{dt}\frac{\gamma(t)(E\!-\!E_{t,h})^2}{(1\!+\!\gamma(t) E)^2 (1\!+\!\gamma(t) E_{t,h})} \!+\! \smallO_L(1).
\end{align}
Since $\gamma(t)$ is an increasing function we see that, quite remarquably, $R_{t,h}$ is negative up to a vanishing term. A similar interpolation technique ensures that the limit $\lim_{L\to \infty} i_{1, h}$ exists. We therefore obtain from \eqref{eq:fcs_fdedn_plus_remainder} that $\lim_{L\to \infty}i_{1,h} \!\leq \!i^{\rm RS}(E;\Delta) \!+\! i_{0,h} \!-\! i_{0,0}$ for a.e $h$ (note from \eqref{eq:int_hamiltonian} that $i_{0,h}$ is independent of $L$ for any $h$ as the signal components are uncorrelated for the denoising models). We can now take the limit $h\!\to \!0$ along a suitable subsequence. It is easy to check $\lim_{h\to 0} i_{0, h} \!=\! i_{0,0}$. Also $\lim_{h\to 0}\lim_{L\to \infty}i_{1,h} \!=\! \lim_{L\to \infty} i_{1,0}$
because concavity and continuity of $i_{1,h}$ and existence of the $L\!\to\! \infty$ limit imply that $\lim_{L\to \infty} i_{1,h}$ is also a concave and continuous function of $h$. We conclude 
$\lim_{L\to\infty}i_{1,0}\! \leq\! i^{\rm RS}(E;\Delta)$, that is equivalent to theorem~\ref{th:upperbound}.
%
%
\subsection{Invariance of the mutual information: proof idea} \label{sec:mutualInfoSCmodel}
In order to prove lemma~\ref{lemma:modelsEquiv}, we compare three models: the \emph{decoupled} $w\!=\!0$ model, 
the SC $0\!<\!w\!<\!(\Gamma\!-\!1)/2$ model and the \emph{homogeneous} $w\!=\!(\Gamma\!-\!1)/2$ model. In all cases, a periodic matrix (fig.~\ref{fig:opSpCoupling}, left) is used. Comparing them directly is rather difficult. Instead, we compare the MI change due to weak perturbations of a ``base'' model where a given (periodic) base measurement matrix is used. We start comparing the homogeneous and coupled models by showing that weakly modifying the base model, in a way that makes it ``closer'' to the homogeneous model than to the coupled one, cannot decrease the MI (up to vanishing terms). This allows to get inequalities between the MI of the two models.

This idea is formalized through the introduction of the \emph{$r$-ensembles} that we describe now. 
Recall that the periodic SC matrices are decomposed in $\Gamma\times\Gamma$ blocks (see fig.~\ref{fig:opSpCoupling}). Focus only on the 
{\it block-row} decomposition.
Consider a virtual 
{\it thinner} decomposition into ``sub-block-rows'': each of the $\Gamma$ block-rows is 
virtually decomposed in $M/(\Gamma L^{u})$ sub-block-rows with $L^u$ lines. The total number of such
sub-block-rows is $\tau\! \defeq\! ML^{-u}$. Let $r\!\in\! \{0,\dots,\tau\}$ and define a periodic SC  
matrix $\bm{\phi}_r$ as follows: $\tau - r$ of 
its sub-block-rows have a coupling window $w_0\!=\!w$ and the remaining $r$ ones have a coupling $w_1\!=\!(\Gamma\!-\!1)/2$. 
This defines the $r$-ensembles. Note that for $r\!=\!0$ we get the usual periodic SC ensemble with window $w$ and for $r\!=\!\tau$ we get the 
homogeneous ensemble with $w\!=\!(\Gamma\!-\!1)/2$. Let $i_r$ denote the MI of the $r$-ensemble. Our goal is to compare $i_{r}$ and $i_{r+1}$
through an appropriate interpolation. This is done by conditionning on a
{\it base matrix} $\bm{\phi}_r^*$ obtained by selecting uniformly 
at random a sub-block-row $s$ among the $\tau\!-\!r$ sub-blocks that have a coupling window $w_0$ and 
removing it from $\bm{\phi}_r$. 
The $r$-ensemble is obtained by further averaging w.r.t the matrix elements of a sub-block which is added back with same index $s$ and coupling window $w_0$. The average over $s$ is also carried out. Similarly, the ($r\!+\!1$)-ensemble is obtained by further averaging 
over a sub-block of index $s$ and
coupling window $w_1$.
With this procedure we are able to carry out the interpolation bounds and show that for a.e $h$,
\be \label{eq:beforeLtoInf}
i_r \le i_{r+1} + L^{u-1}\smallO_L(1) + \mathcal{O}(L^{u-1+u-1/8}).
\ee
The first correction in the inequality comes from concentrations needed in the proof (similar to the one in paragraph \ref{sec:ymmse} below) while 
the second is due to the slight difference between the mmse's of the $r$ and ($r\!+\! 1$)-ensembles. 
Inequality \eqref{eq:beforeLtoInf} implies that the MI difference $i^{\rm per}_{\Gamma, w} - i^{\rm per}_{\Gamma,  (\Gamma-1)/2} = i_0 - i_{\tau}$ is equal to
\begin{align}
\sum_{r=0}^{\tau-1}(i_r - i_{r+1}) \le \tau (L^{u-1}\smallO_L(1) + \mathcal{O}(L^{u-1+u-1/8})). \label{eq:beforeLtoInf_1}
\end{align}
Choosing $u$ small enough, say $u\!=\!1/16$, and since we have that $\tau\!=ML^{-u}=\!\mathcal{O}(L^{1-u})$, we finally reach that $\lim_{L\to\infty}i^{\rm per}_{\Gamma, (\Gamma-1)/2}\ge \lim_{L\to\infty}i^{\rm per}_{\Gamma, w}$.

Using the same strategy with $w_0\!=\!w$ and $w_1\!=\!0$, we show an inequality between the MI of the SC and decoupled models which leads with the previous one to 
\be
\lim_{L\to\infty} i^{\rm per}_{\Gamma, (\Gamma-1)/2} \! \geq \!\lim_{L\to\infty} i^{\rm per}_{\Gamma, w}\!\geq\! \lim_{L\to\infty} i^{\rm per}_{\Gamma,0}.
\ee
Since the two extreme limits are equal by construction to $\lim_{L\to\infty} i^{\rm cs}$, we obtain the first part of lemma~\ref{lemma:modelsEquiv}. Note that obtaining this inequality is the reason for the introduction of the SC periodic model rather than studying directly the SC seeded model for which threshold saturation holds; the periodicity makes this analysis much simpler.

We conclude by the similar statement for the seeded model. It is not hard to show that the MI difference between those of the periodic and seeded SC models is $\mathcal{O}(w/\Gamma)$, thus vanishing when $\Gamma\!\to\!\infty$ for any fixed $w$. As a consequence, for the seeded model with fixed $w$ we have $\lim_{\Gamma\to\infty}\lim_{L\to\infty}i^{\rm seed}_{\Gamma, w}\!=\! \lim_{L\to\infty} i^{\rm cs}$ as well, proving the second part of lemma~\ref{lemma:modelsEquiv}.
\subsection{Computing ${\rm ymmse}_{t,h}$} 
\label{sec:ymmse}
Let us now prove lemma~\ref{lemma:MSEequivalence}. A direct application of the Nishimori identity (remark~\ref{remark:Nishi}) brings $2\EE[\langle[\bm{\Phi}\bar\bX]_\mu\rangle_{t,h}^2]\!=\!\EE[\langle[\bm{\Phi}\bar\bX]_\mu^2\rangle_{t,h}]$. Using this, \eqref{eq:A_twoterms} and the first equality of \eqref{eq:secondTerm_fint_A}, we obtain that ${\rm ymmse}_{t,h}\!\defeq\!\sum_{\mu}\mathbb{E}[\langle [\bm{\Phi}\bar \bX]_\mu \rangle_{t,h}^2]/M$ is also equal to
\begin{align}
&\frac{1}{M}\!\sum_{\mu=1}^M \!\mathbb{E}\Big[\Big\langle [\bm{\Phi}\bar \bX]_\mu^2 \!- \!\frac{[\bm{\Phi}\bar \bX]_\mu Z_{\mu}}{\sqrt{\gamma}}\, \Big\rangle_{t,h}\Big]\!=\! \frac{1}{2M}\! \sum_{\mu=1}^M\! \mathbb{E}[\langle [\bm{\Phi}\bar \bX]_\mu^2 \rangle_{t,h}] \nonumber\\
&\Rightarrow {\rm ymmse}_{t,h} \!= \!\frac{1}{\sqrt{\gamma}M}\!\sum_{\mu=1}^M\EE[Z_{\mu}\langle [\bm{\Phi}\bar \bX]_\mu\rangle_{t,h}]. \label{eq:ident_ymmse}
\end{align}
%
Define $U_\mu \!\defeq\! \sqrt{\gamma} \,[\bm{\Phi} \bar \bX]_\mu \!-\! Z_{\mu}$. An integration by part w.r.t $\phi_{\mu i}\!\sim\!\mathcal{N}(0,1/L)$ of \eqref{eq:ident_ymmse} brings that
\begin{align}
{\rm ymmse}_{t,h}\!\!=\!\!\frac{1}{ML} \!\!\sum_{\mu,i=1}^{M,N}\!\!\EE[Z_{\mu} \langle U_\mu {\bar  X}_i\rangle_{t,h} \langle {\bar  X}_i\rangle_{t,h}\!-\!Z_{\mu}\langle U_\mu {\bar X}^2_i\rangle_{t,h}].\nonumber
\end{align}
The Nishimori identity allows to write from the last expression that ${\rm ymmse}_{t,h}$ equals
\begin{align}
&\!\frac{1}{ML}\!\!\sum_{\mu, i=1}^{M,N}\!\! \Big(\!-\! \sqrt{\gamma}\, \EE[Z_{\mu} S_i \langle [\bm{\Phi}{\bar \bX}]_\mu {\bar X}_i\rangle_{t,h}] \!-\!\EE[Z_{\mu} \langle U_\mu {\bar X}^2_i\rangle_{t,h}] \nonumber\\
&+\! \EE[Z_{\mu}^2 S_i \langle{\bar X}_i\rangle_{t,h}]\Big)\!= \! \sum_{\mu, i=1}^{M,N} \!\! \Big(\EE[Z_{\mu}^2 \langle \bar X_i^2\rangle_{t,h}] \!+\! \EE[Z_{\mu}^2 S_i \langle{\bar X}_i \rangle_{t,h} ] \nonumber\\
&-\! \sqrt{\gamma}\,\EE[Z_{\mu} \langle [\bm{\Phi} \bar \bX]_\mu X_i {\bar X}_i\rangle_{t,h}] \Big) \frac{1}{ML}\!=\! {\cal Y}_1 - {\cal Y}_2, \label{eq:allpieces} 
%
%
\end{align}
together with 
\begin{align}
{\cal Y}_1\!&=\!\EE\Big[\frac{1}{M}\sum_{\mu=1}^M\!Z_{\mu}^2 \frac{1}{L}\sum_{i=1}^N \langle X_i \bar X_i\rangle_{t,h}\Big],\nonumber\\
{\cal Y}_2\!&=\! \sqrt{\gamma}\, \EE\Big[\frac{1}{M}\sum_{\mu=1}^M\! Z_{\mu}\Big\langle [\bm{\Phi} \bar \bX]_\mu  \frac{1}{L}\sum_{i=1}^N  X_i \bar X_i\Big\rangle_{t,h}\Big].
\end{align}
By the law of large numbers, $\sum_\mu z_{\mu}^2/M \!=\! 1 \!+\!\smallO_L(1)$ almost surely as $L\!\to\!\infty$ so that using the Nishimori identity, we reach ${\cal Y}_1\!=\!E_{t,h}\!+\!\smallO_L(1)$.
%
%
Using similar concentration proofs as~\cite{KoradaMacris_CDMA} (and this is actually the point where the perturbation of the interpolated model becomes fundamental), one can show for the second term that for a.e $h$
\begin{align*}
{\cal Y}_2 \!&=\!   \EE\Big[\sum_{i=1}^N\!\! \frac{\langle X_i \bar X_i\rangle_{t,h}}{L}\Big]\sqrt{\gamma}\, \EE\Big[\sum_{\mu=1}^M \frac{Z_{\mu}}{M}\langle [\bm{\Phi} \bar \bX]_\mu\rangle_{t,h}\Big] \!+\!\smallO_L(1) \nonumber\\
&= \!E_{t,h}\sqrt{\gamma}\, \frac{1}{M}\sum_{\mu=1}^M\EE[Z_{\mu}\langle [\bm{\Phi} \bar \bX]_\mu\rangle_{t,h}]\! +\!\smallO_L(1),
\end{align*}
where we used the Nishimori identity to identify $E_{t,h}$ in the second equality. From this and \eqref{eq:ident_ymmse} we recognize ${\cal Y}_2 \!=\!E_{t,h}  \gamma(t){\rm ymmse}_{t,h}\!+\! \smallO_L(1)$. Putting all pieces together, we get from \eqref{eq:allpieces} that ${\rm ymmse}_{t,h}\!=\! E_{t,h}\! -\! E_{t,h}  \gamma(t){\rm ymmse}_{t,h} \!+\! \smallO_L(1)$ for a.e $h$, which leads to lemma~\ref{lemma:MSEequivalence}.
%
%
\section*{Acknowledgments}
J.B and M.D acknowledge funding from the FNS (grant
200021-156672). F.K thank the Simons Institute in Berkeley for its hospitality and acknowledge funding from the EU (FP/2007-2013/ERC grant agreement
307087-SPARCS).
\bibliographystyle{IEEEtran}
\bibliography{refs}
\end{document}